\documentclass[aps,pre,twocolumn,a4paper,byrevtex,superscriptaddress,showpacs,showkeys,longbibliography]{revtex4-1}
\usepackage{microtype}
\usepackage{graphicx}
\usepackage{amsmath}
\usepackage{amssymb}
\usepackage{color}
\definecolor{myblue}{rgb}{0.153,0.322,0.706}
\usepackage[colorlinks,linkcolor=myblue,urlcolor=myblue,citecolor=myblue]{hyperref}

\newcommand{\be}{\begin{equation}}
\newcommand{\ee}{\end{equation}}
\newcommand{\ra}{\rightarrow}
\newcommand{\reals}{\mathbb{R}}
\newcommand{\cL}{\mathcal{L}}

\newcommand{\inv}{\text{inv}}
\newcommand{\hth}{\hat\theta}
\newcommand{\var}{\textrm{var}}
\newcommand{\ent}{\textrm{ent}}
\newcommand{\qap}{\textrm{aux}}
\newcommand{\md}{m}
\newcommand{\tI}{\tilde I}
\newcommand{\tL}{\tilde\lambda}
\newcommand{\tF}{\tilde F}
\newcommand{\bu}{\nu}
\newcommand{\cI}{\mathcal{I}}

\begin{document}
\title{Current large deviations for driven periodic diffusions}

\author{Pelerine Tsobgni Nyawo}
\email{tsobgnipelerine@gmail.com}
\affiliation{\mbox{Institute of Theoretical Physics, Department of Physics, University of Stellenbosch, Stellenbosch 7600, South Africa}}

\author{Hugo Touchette}
\email{htouchette@sun.ac.za, htouchet@alum.mit.edu}
\affiliation{National Institute for Theoretical Physics (NITheP), Stellenbosch 7600, South Africa}
\affiliation{\mbox{Institute of Theoretical Physics, Department of Physics, University of Stellenbosch, Stellenbosch 7600, South Africa}}

\begin{abstract}
We study the large deviations of the time-integrated current for a driven diffusion on the circle, often used as a model of nonequilibrium systems. We obtain the large deviation functions describing the current fluctuations using a Fourier-Bloch decomposition of the so-called tilted generator and also construct from this decomposition the effective (biased, auxiliary or driven) Markov process describing the diffusion as current fluctuations are observed in time. This effective process provides a clear physical explanation of the various fluctuation regimes observed. It is used here to obtain an upper bound on the current large deviation function, which we compare to a recently-derived entropic bound, and to study the low-noise limit of large deviations.
\end{abstract}

\date{\today}


\maketitle

\section{Introduction}

We study in this paper the driven diffusion on the circle defined by the following stochastic differential equation (SDE):
\be
d\theta_t= [\gamma-V'(\theta_t)]dt+\sigma dW_t,
\label{eqsde1}
\ee
where $\theta_t\in [0,2\pi)$, $V(\theta)$ is a periodic potential taken to be
\be
V(\theta)=V_0 \cos\theta,
\label{eqcospot1}
\ee
$\gamma\in\reals$ is a constant driving frequency, and $W_t$ is a Brownian motion multiplied by the noise intensity $\sigma\geq 0$. This SDE represents one of the simplest nonequilibrium system violating detailed balance for $\gamma\neq 0$ and has played, as such, an important role in the development and illustration of recent results about nonequilibrium response \cite{reimann2001,reimann2002b,blickle2007,seifert2010}, entropy production \cite{speck2007a,mehl2008,speck2012}, and large deviations in the long-time \cite{maes2008,alexander1997,nemoto2011b,nemoto2011,saito2015} or low-noise \cite{freidlin1984,graham1995,faggionato2012} regime. It is also used as a model of Josephson junctions subjected to thermal noise \cite{kautz1988,kautz1996,risken1996}, Brownian ratchets \cite{reimann2002}, and manipulated Brownian particles \cite{gomez2009,ciliberto2010,gomez-solano2011}, among other systems (see \cite{reimann2001}), and is thus an ideal experimental testbed for the physics of nonequilibrium systems.

In this paper, we use large deviation theory to study the fluctuations of a natural observable of the driven diffusion, its mean velocity, defined formally as
\be
J_T=\frac{1}{T}\int_0^T \dot \theta_t\, dt.
\label{eqcurr1}
\ee
Previous works have looked at the large deviations of this quantity \cite{nemoto2011b,nemoto2011}, which also represents a time-integrated, fluctuating current for the diffusion, as well as the large deviations of the entropy production \cite{speck2007a,mehl2008,speck2012}, which is linearly related to $J_T$. Our goal here is to complete these studies by investigating the large deviation functions characterizing the fluctuations of $J_T$ in all noise regimes, and by constructing the \textit{auxiliary process}, also known as the \textit{biased} or \textit{driven process}, describing the diffusion conditionally on observing a current fluctuation $J_T=j$ far from the mean current $\langle J_T\rangle$. This process is physically important as it describes, by means of a modified stochastic process, how fluctuations of the current or any time-integrated observable in general are created in time \cite{evans2004,evans2005a,jack2010b,chetrite2013,chetrite2015,chetrite2014}.  

This effective description of fluctuations was illustrated recently in the context of interacting particle systems \cite{popkov2010,popkov2011,harris2013b,jack2014,hirschberg2015}, diffusions \cite{simha2008,knezevic2014,angeletti2015}, and quantum systems \cite{garrahan2010,garrahan2011,ates2012,genway2012,hickey2012}. For the SDE (\ref{eqsde1}), preliminary results \cite{chetrite2013} have shown that the auxiliary process modifies not only the driving $\gamma$, which is a natural way to increase or decrease the current, but also the potential $V(\theta)$ in a non-local and nonlinear way. Here, we complete these results by constructing the auxiliary process for a wider range of parameters and by relating it to the different fluctuation regimes seen at the level of the large deviation functions. We also study fluctuation symmetries for $J_T$, related to the so-called fluctuation relation for the entropy production \cite{speck2007a,mehl2008,speck2012}, and demonstrate an entropic bound for the rate function recently derived in \cite{gingrich2016} (see also \cite{pietzonka2015}). 

The results that we obtain show a rich trade-off between modifying $\gamma$ and $V(\theta)$ to reach low or high current fluctuations, yielding many physical insights about how these fluctuations arise in time. This is particularly useful for understanding the low-noise limit of large deviations, studied within the so-called Freidlin-Wentzell theory \cite{freidlin1984} (see also \cite{graham1989}) or the macroscopic fluctuation theory \cite{bertini2002,bertini2006,bertini2007} in terms of most probable paths or instantons minimizing a given stochastic action. We show here how to use the deterministic limit of the auxiliary process as an alternative way to recover these instantons. Using this technique, we are able to clarify certain properties of the rate function for the circle diffusion related to a dynamical phase transition.

\section{Current large deviations}
\label{seccurrldt}

We briefly explain in this section the large deviation formalism used to describe the probability distribution of $J_T$ in the long-time limit and how the auxiliary process is constructed from spectral elements related to the large deviations of $J_T$. For background material on large deviations, see~\cite{dembo1998,hollander2000,touchette2009}.

\subsection{Large deviation principle}

The paths of pure diffusions are nowhere differentiable, as is well known, so the expression of the current shown in (\ref{eqcurr1}) is only a formal expression that we replace mathematically by the stochastic integral
\be
J_T=\frac{1}{T}\int_0^T d\theta_t = \frac{(\theta_T-\theta_0)N_T}{T},
\ee
where $N_T$ is the \textit{winding number}, that is, the net number of turns done by $\theta_t$ through $\theta=0$ (or any other angle) after a time $T$. Alternatively, we can write
\be
J_T=\frac{\theta_T-\theta_0}{T}
\ee
by considering $\theta_t$ to be a multivalued angle taking values in $\reals$ instead of $[0,2\pi)$. Without loss in generality, we choose $\theta_0=0$ as the initial angle.

In the infinite-time or ergodic limit, $J_T$ is known to converge to the average speed $\langle J_T\rangle$, given for the driven periodic diffusion by the expectation
\be
\langle J_T\rangle =\langle F(\theta)\rangle
\ee
of the total force
\be
F(\theta)=\gamma-V'(\theta)
\label{eqf1}
\ee 
driving the SDE (\ref{eqsde1}). The exact expression of this expectation, due to Stratonovich, can be found in \cite{reimann2001} (see also the formula (25) in \cite{speck2012}). With this result, we thus have
\be
\lim_{T\ra\infty} J_T = \langle F\rangle
\ee
for almost all paths of the diffusion, which means that, although $J_T$ fluctuates around its mean, it converges almost surely to it as $T\ra\infty$. For this reason, the mean is also called the \emph{concentration point} of $J_T$.

We are interested here in the rare fluctuations of $J_T$ around this concentration point. Following the theory of large deviations (see, e.g., \cite{dembo1998,hollander2000,touchette2009}), the probability of these fluctuations has the general form
\be
P(J_T=j)\approx e^{-TI(j)}
\label{eqldp1}
\ee
in the limit $T\ra\infty$. The approximation sign means that corrections to the exponential term are sub-linear in $T$ in the exponent, which means that the exponential itself is the dominant term of $P(J_T=j)$ at large times. The function $I(j)$ given by the limit
\be
I(j)=\lim_{T\ra\infty}-\frac{1}{T}\ln P(J_T=j)
\ee
is called the \emph{rate function} and is so named because it controls the rate at which the probability $P(J_T=j)$ decays to zero for any $j\neq\langle J_T\rangle$ as $T\ra\infty$. As a result, we have $I(\langle J_T\rangle)=0$ and $I(j)>0$ for any other values of $j$, showing that the fluctuations of $J_T$ away from its mean are exponentially unlikely at large times.

For the model studied here, $I(j)$ is exactly quadratic and equal to 
\be
I(j)=\frac{(j-\gamma)^2}{2\sigma^2}
\label{eqqrft1}
\ee
when $V_0=0$ \cite{maes2008}. For other parameter values, it has a non-trivial shape characterizing non-Gaussian current fluctuations coming from the interplay between the potential $V(\theta)$ and the nonequilibrium drive $\gamma$.

\subsection{Large deviation functions}

Many methods can be used to obtain the rate function $I(j)$; here we use the G\"artner-Ellis Theorem \cite{dembo1998,hollander2000,touchette2009}, which states that $I(j)$ is given by the Legendre-Fenchel transform of the \textit{scaled cumulant generating function} (SCGF) of $J_T$,
\be
\lambda(k)=\lim_{T\ra\infty}\frac{1}{T}\ln \langle e^{TkJ_T}\rangle,
\ee
when the latter function exists and is differentiable for $k\in\reals$. Thus,
\be
I(j)=\sup_{k\in\reals}\{kj-\lambda(k)\}
\label{eqlf1}
\ee
under these conditions.

For time-integrated observables of Markov processes such as $J_T$, $\lambda(k)$ is known to be given by the dominant eigenvalue of a modified linear operator, called the \textit{tilted generator}, which corresponds here to 
\be
\cL_k=F\left(\frac{d}{d\theta} +k\right)+\frac{\sigma^2}{2}\left(\frac{d}{d\theta}+k\right)^2
\label{eqtg1}
\ee
and which acts on periodic functions of $[0,2\pi)$ (see \cite{chetrite2014} for the derivation of this operator). As a result, we write
\be
\cL_k r_k(\theta)=\lambda(k)r_k(\theta),
\label{eqspectral1}
\ee
where $\lambda(k)$ is the dominant eigenvalue of $\cL_k$ and $r_k(\theta)$ is its corresponding (periodic) eigenfunction. 

For $k=0$, $\cL_0=L$ is simply the generator of the SDE (\ref{eqsde1}) having the trivial eigenfunction $r_0(\theta)=1$, which is conjugated to the stationary density $\rho^\inv(\theta)$ solving the Fokker-Planck equation
\be
L^\dagger \rho^\inv=0,
\label{eqfp1}
\ee
where $L^\dag$ is the adjoint of $L$. The stationary density $\rho^\inv(\theta)$ is unique, since the process is defined on a compact space, and is given by an explicit formula for all parameter values; see Chap.\ 11 of \cite{risken1996}. The compactness of the space also implies that the spectrum of $\cL_k$ is discrete. It is real in the equilibrium case ($\gamma=0$) and is composed otherwise of complex-conjugate pairs of eigenvalues, except for the dominant eigenvalue $\lambda(k)$ which is always real.

The non-hermitian spectral problem (\ref{eqspectral1}) has no known solution, except for $k=0$ and for $V_0=0$. However, its general solution can be constructed easily, following \cite{mehl2008}, by means of a Fourier-Bloch decomposition of the eigenfunction
\be
r_k(\theta)=\sum_{n\in\mathbb{Z}} c_n\, e^{in\theta}
\ee
and the potential derivative
\be
V'(\theta)=\sum_{n\in\mathbb{Z}} v_n\, e^{in\theta}.
\ee
Substituting these expansions in (\ref{eqspectral1}) yields a recurrence relation for the coefficients $c_n$, which reduces for the cosine potential (\ref{eqcospot1}) to
\be
b_n^-c_{n-1}+[a_n-\lambda(k)]c_n+b^+_n c_{n+1}=0,
\ee
where
\begin{eqnarray}
a_n &=& in (\gamma+\sigma^2k)-\frac{\sigma^2n}{2}+k\gamma+\frac{k^2\sigma^2}{2}\nonumber\\
b_n^\pm &=& \frac{V_0}{2} (\pm ik-1\mp n).
\end{eqnarray}

To solve this tri-diagonal system, we naturally truncate $n$ to some discrete range $[\![ -M,M ]\!]$, yielding a system of $2M+1$ linearly independent equations, which are  solved numerically to find the coefficients $c_n$ of $r_k(\theta)$, the SCGF $\lambda(k)$, and the rate function $I(j)$ by Legendre-transforming $\lambda(k)$. We present the results of these calculations for various parameters in the next section. In all cases, we have checked that the results converge for $M$ large enough and only present the largest $M$ used, which is typically between $M=5$ and $M=30$ modes depending mostly on the ratio $\sigma/V_0$. In general, the smaller $\sigma/V_0$ is, the larger $M$ must be chosen \footnote{\label{fn1}In practice, we are able to use Mathematica (version 10) to study noise intensities as small as $\sigma=0.35$ for $V_0=1$, $\gamma\approx V_0$ and $k$ in the relevant range. For smaller values of $\sigma$, the linear system to solve becomes unstable.}.

For the rest of the paper, it is useful to note that the large deviation functions $\lambda(k)$ and $I(j)$ can also be obtained in a very different way via optimization problems derived in \cite{chetrite2015} (see also \cite{jack2015,nemoto2011b,nemoto2011}). For the SCGF, the optimization to solve is
\be
\lambda(k)=\inf_u \{ k\langle u\rangle_u- K(u)\}
\label{eqopt1}
\ee
where 
\be
K(u)=\frac{1}{2\sigma^2}\int_0^{2\pi} [u(\theta)-F(\theta)]^2 \rho^\inv_u(\theta)\, d\theta
\ee
and
\be
\langle u\rangle_u=\int_0^{2\pi} u(\theta) \rho^\inv_u(\theta) d\theta
\ee
is the average current calculated with respect to the stationary density $\rho^\inv_u(\theta)$ of a diffusion with total force $u(\theta)$, that is, the stationary density solving the Fokker-Planck equation (\ref{eqfp1}) with $F(\theta)$ replaced by $u(\theta)$. By Legendre duality, the rate function is then obtained by solving the constrained optimization
\be
I(j)=\inf_{u: \langle u\rangle_u=j} K(u).
\label{eqopt2}
\ee
In both cases, the optimization is over all continuous and periodic functions $u(\theta)$.

Similar optimizations were considered in \cite{alexander1997,nemoto2011b,nemoto2011} as a way to study the large deviations of the ring model. They are difficult to solve in general, but can be expanded in Fourier-Bloch modes at the level of $u(\theta)$ to give what is essentially the eigenfunction $r_k(\theta)$ constructed above. In some cases, exact solutions can be found, as will be discussed in the next section, in addition to approximate solutions for $u(\theta)$, which yield useful approximations and bounds for $\lambda(k)$ and $I(j)$.

\subsection{Effective fluctuation process}

The auxiliary or driven process mentioned in the introduction is constructed from the dominant eigenfunction $r_k(\theta)$ as the new diffusion $\hth_t$ given by the SDE
\be
d\hth_t=F_k(\hth)dt+\sigma dW_t,
\label{eqauxp1}
\ee
which involves the same noise as $\theta_t$ but a modified force
\be
F_k(\theta)=F(\theta)+\sigma^2\left(k+\frac{d}{d\theta}\ln r_k(\theta)\right)
\label{eqmodf1}
\ee
compared to the force $F(\theta)$ of $\theta_t$ \cite{chetrite2013}. The idea again behind this process is to understand how the original process $\theta_t$ reaches a current fluctuation $J_T=j$ after a long time $T$ \cite{chetrite2013}. In mathematical terms, this means that we must condition $\theta_t$ on the event $J_T=j$ and infer from this conditioning a new Markov process that  describes the set of ``constrained'' paths of $\theta_t$ such that $J_T=j$ \cite{chetrite2014}. 

\begin{figure*}[t]
\centering
\resizebox{\textwidth}{!}{\includegraphics{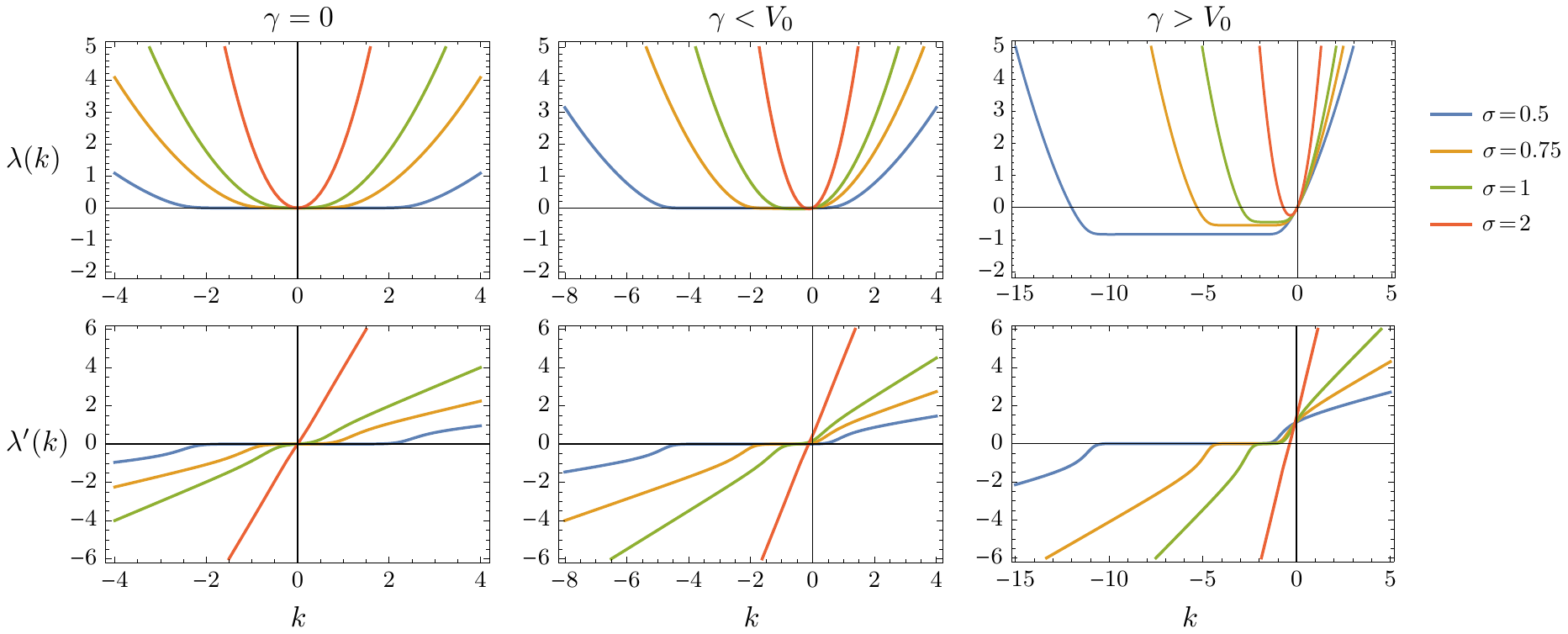}}
\caption{(Color online) Top row: SCGF for different values of $\sigma$. Bottom row: Derivative of the SCGF. Parameters: $V_0=1$, $\gamma=0$ (left column), $\gamma=0.5$ (middle column), $\gamma=1.5$ (right column).}
\label{figld1}
\end{figure*}

The auxiliary process is that Markov process. To be more precise, it is the process that is \textit{equivalent to the conditioned process in the long-time limit}, in the same way that the canonical ensemble is equivalent to the microcanonical ensemble in the infinite-volume limit \cite{chetrite2014}. In fact, equivalence is achieved similarly to equilibrium by relating the constant current $j$ of the conditioned (microcanonical) process to the ``temperature'' $k$ of the auxiliary (canonical) process according to \cite{chetrite2013}
\be
I'(j)=k
\ee 
or, equivalently,
\be
\lambda'(k)=j.
\ee 
From this, it is natural to interpret $\hth_t$ as the effective process that ``creates'' the fluctuation $J_T=j$, just as the canonical ensemble ``creates'' in the thermodynamic limit a microcanonical ensemble with fixed energy. Naturally, $F_{k=0}=F$ since $\lambda'(0)=\langle J_T\rangle$.

It is worth emphasising that the constraint $J_T=j$ is not satisfied at all times in the auxiliary process. What we have again is a long-time or ergodic equivalence implying that $J_T\ra j$ for this process as $T\ra\infty$. As a result, the rare event that is $J_T=j$ for $\theta_t$ is transformed into a typical event for $\hth_t$, which is useful for simulations \cite{chetrite2015}. From the point of view of control theory, it can be shown that this change of process minimizes $K(u)$ above, so that the optimal $u(\theta)$ in (\ref{eqopt1}) or (\ref{eqopt2}) is actually $F_k(\theta)$ \cite{chetrite2015}. This explains why the Fourier-Bloch solution for $u(\theta)$ is equivalent, as mentioned, to the Fourier-Bloch solution for $r_k(\theta)$. The two are related by (\ref{eqmodf1}).

\section{Results}
\label{secres}

We present in this section the results of the Fourier-Bloch solution of the large deviation functions and the auxiliary process. Some of these results are related to the large deviations of the mean entropy production $\Sigma_T$ \cite{speck2007a,mehl2008,speck2012}, which is linearly related to the current $J_T$ according to
\be
\Sigma_T= \frac{2\gamma}{\sigma^2} J_T-\frac{2}{\sigma^2 T}[V(\theta_T)-V(\theta_0)],
\label{eqep1}
\ee
while others have appeared in the context of variational approaches to large deviations \cite{alexander1997,nemoto2011b,nemoto2011}. Consequently, our discussion of the large deviation functions, which are known to some extent, will be brief. Our main goal, as mentioned before, is to explain with the auxiliary process how the different fluctuation regimes inferred from these functions arise physically. 

To understand these results, it is important to note that the noiseless ($\sigma=0$) dynamics undergoes a bifurcation between a \textit{fixed-point solution} for $|\gamma|\leq V_0$, where $\dot\theta_t\ra 0$ after some transient time so that $J_T\ra 0$, and a \textit{running solution} for $|\gamma|> V_0$, where $\theta_t$ rotates in such a way that
\be
\lim_{T\ra\infty} J_T = \sqrt{\gamma^2-V_0^2}
\label{eqdetj1}
\ee
for $\gamma>V_0$ \cite{strogatz1994}. This bifurcation is ``rounded'' by the noise, but it still determines much of the different fluctuation regimes discussed next.

\subsection{Current fluctuations}

\begin{figure*}[t]
\resizebox{\textwidth}{!}{\includegraphics{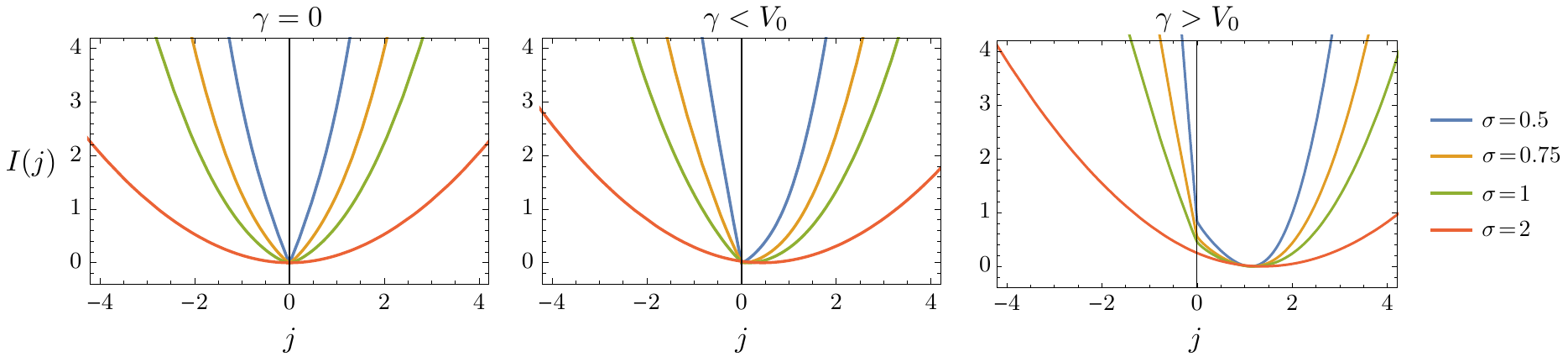}}
\caption{(Color online) Rate function for different values of $\sigma$. Parameters: $V_0=1$, $\gamma=0$ (left column), $\gamma=0.5$ (middle column), $\gamma=1.5$ (right column).}
\label{figld2}
\end{figure*}

We show in Fig.~\ref{figld1} the plot of $\lambda(k)$ as a function of $k$ for different values of $\gamma$ and $\sigma$, together with the plot of its derivative for the same parameters. From the results, we can see that the current fluctuations are essentially Gaussian at high noise (i..e, large $\sigma$ relative to $V_0$), since $\lambda(k)$ is then a parabola (with linear slope), which implies that the rate function $I(j)$, obtained by the Legendre transform (\ref{eqlf1}), is also a parabola, as seen in Fig.~\ref{figld2}. For low noise, $\lambda(k)$ develops instead a flat plateau giving rise, by Legendre transform, to a ``kink'' in $I(j)$ around $j=0$, indicating a  non-Gaussian crossover between negative and positive fluctuations. Far away from $j=0$, the fluctuations become Gaussian again, as can be seen by the fact that $\lambda'(k)$ is linear away from its plateau. 

\begin{figure}[t]
\includegraphics{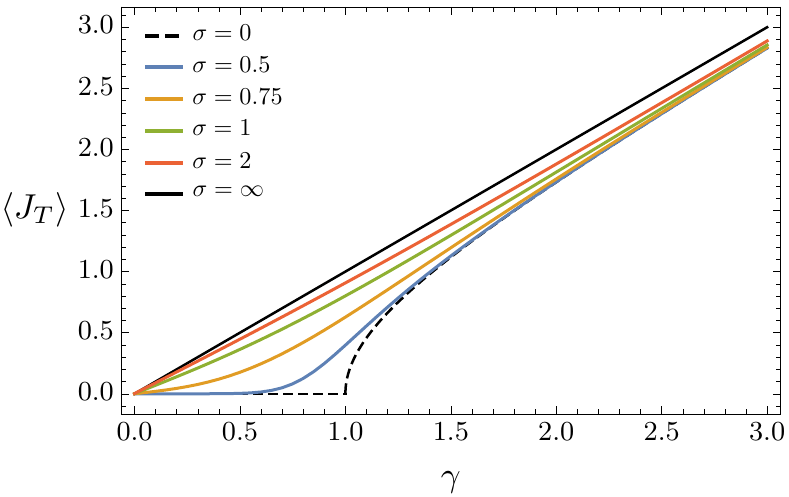}
\caption{(Color online) Mean current as a function of $\gamma$ for $V_0=1$. The mean current is given by $\lambda'(0)$ or, equivalently, by the zero of $I(j)$.}
\label{figmean1}
\end{figure}

This picture remains more or less the same when the nonequilibrium drive $\gamma$ is increased; all that changes is the value of $\lambda'(0)$ which determines the zero of $I(j)$ and thus the mean current, shown in Fig.~\ref{figmean1}. From this plot, we see that $\langle J_T\rangle$ is essentially zero in the fixed-point regime when $\sigma$ is small, and grows according to (\ref{eqdetj1}) in the running regime. For large $\sigma$, we find instead $\langle J_T\rangle \approx \gamma$. In each case, the kink of the rate function $I(j)$ remains at $j=0$, despite the zero of $I(j)$ moving with $\gamma$, and becomes more pronounced as $\sigma\ra 0$. 


This kink was already reported for the entropy production \cite{speck2007a,mehl2008,speck2012} and is akin to dynamical phase transitions seen in the activity or current fluctuations of particle models \cite{garrahan2007,garrahan2009,hedges2009,hooyberghs2010,dickson2011,espigares2013} and disordered random walks \cite{gingrich2014}. For the ring model, there is no dynamical phase transition properly speaking because $\lambda(k)$ is not exactly flat in the plateaus: it only grows very slowly from a central minimum, as can be verified by zooming in the plateau. This implies that $I(j)$ has a rounded kink at $j=0$ with continuous derivative instead of an actual cusp with discontinuous derivative. In general, for $\lambda(k)$ and $I(j)$ to have non-analytic points, determining either a phase transition in the mean of $J_T$ or a dynamical phase transitions in its fluctuations, there needs to be a thermodynamic or scaling limit, such as the noiseless limit studied in Sec.~\ref{seclownoise} \footnote{It is widely believed, though not proved rigorously, that large deviation functions of time-additive observables are always analytic in the long-time limit for Markov processes with compact state space.}.

\begin{figure}[t]
\includegraphics{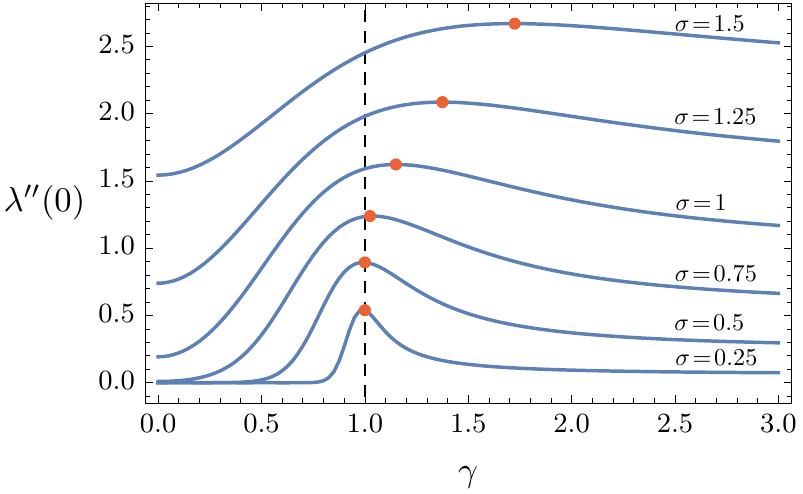}
\caption{(Color online) Asymptotic current variance as a function of $\gamma$ for $V_0=1$ and different values of $\sigma$. The dots show the maximum of each curve.}
\label{figvar1}
\end{figure}

Another crossover in the fluctuations can be seen at the level of $\lambda''(0)$, which determines the asymptotic variance of $J_T$ according to 
\be
\lambda''(0)=\lim_{T\ra\infty} T\,\var(J_T).
\ee
The plot of this quantity in Fig.~\ref{figvar1} shows that the current fluctuations around the mean are enhanced near the bifurcation point $\gamma=V_0$, especially at low noise. This is commonly observed in noisy dynamics undergoing bifurcations or phase transitions. A similar crossover, referred to as a ``giant acceleration'' or ``giant response'', is observed for the long-time variance of $\theta_t$, which determines the diffusion coefficient \cite{reimann2001,reimann2002b,blickle2007}.


\subsection{Auxiliary process}

\begin{figure*}[t]
\centering
\resizebox{0.97\textwidth}{!}{\includegraphics{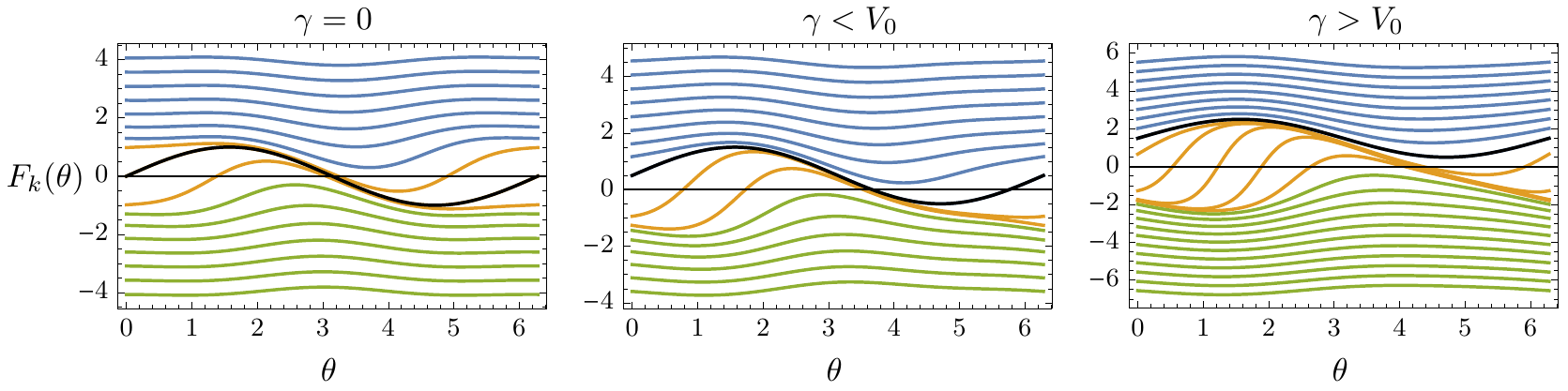}}
\caption{(Color online) Effective force $F_k(\theta)$ of the auxiliary process for the same values of $\gamma$ as in Fig.~\ref{figld1}. The curves in each plot show $F_k$ for different values of $k$ taken, in spacing of 0.5, in the ranges of $\lambda(k)$ used in Fig.~\ref{figld1}. Black curve: $F_0=F$; blue curves: $F_k$ associated with positive current fluctuations; green curves: $F_k$ associated with negative current fluctuations; orange curves: $F_k$ with stable fixed-points (nodes) associated with near-zero current fluctuations. Other parameters: $V_0=1$, $\sigma=1$.}
\label{figdp1}
\end{figure*}

The two mean features of the current fluctuations just discussed, namely, the rounded kink at $j=0$ and the large Gaussian fluctuations away from that kink, can be understood in a clear and physical way by analyzing the auxiliary process. To that end, we show in Fig.~\ref{figdp1} the modified force $F_k(\theta)$ of this process, given by (\ref{eqmodf1}), for the same values of $\gamma$ as in Fig.~\ref{figld1}. We also show $F_k(\theta)$ in each plot for different values of $k$ in the ranges used in Fig.~\ref{figld1}.

From the left plot of Fig.~\ref{figdp1}, corresponding to $\gamma=0$, we clearly see that the effective process associated with large values of $|k|$ (blue and green curves), which encode the large current fluctuations, is to a good approximation a simple diffusion with constant drift given by $F_k(\theta)=\sigma^2 k$. Using this in the variational representation (\ref{eqopt1}), we then find
\be
\lambda(k)\approx \frac{\sigma^2 k^2}{2}
\label{eqlk1}
\ee
to leading order in $k$, which yields from either (\ref{eqlf1}) or (\ref{eqopt2}),
\be
I(j)\approx \frac{j^2}{2\sigma^2}.
\ee
For large fluctuations, the diffusion therefore acts as if there were no potential: the natural drive $\gamma$ is simply modified to create a larger or smaller current. This is  the most efficient way of creating large current values and leads according to (\ref{eqopt2}) to Gaussian fluctuations, as in the case $V_0=0$. 

For the fluctuations close to $j=0$, corresponding to low values of $k$ (orange curves), the effective force is not so trivial. Instead of being globally shifted upwards or downwards, it is modified locally away from the fixed point $\theta=\pi$ in such a way as to bring the unstable fixed point, corresponding to the other node of $F_k(\theta)$, closer to $\pi$. This has the effect of lowering the potential barrier associated with $F_k$ on the left or right of the fixed point, depending on the sign of $k$, thereby creating a small positive or negative current. This is an optimal strategy for creating a current according to (\ref{eqopt2}), as $(u-F)^2$ is minimized around the fixed point where $\theta_t$ spends most of its time and where $\rho^\inv_{u}(\theta)$ is therefore maximum. The fluctuations in this case are non-Gaussian because of the non-constant change from $F(\theta)$ to $F_k(\theta)$.

This dichotomous picture between large fluctuations created by an effective constant drive, on the one hand, and small fluctuations created by lowering the potential barrier, on the other, is consistent physically with what we see for the mean current and also explains the results obtained for $\gamma\neq 0$. In this case, the auxiliary force is shifted to $F_k(\theta)=\gamma +\sigma^2 k$ for large $|k|$, which yields the Gaussian approximation
\be
I(j)\approx\frac{(j-\gamma)^2}{2\sigma^2}
\ee
for the large current fluctuations. For $\gamma>0$, the range of $k$ where $F_k(\theta)$ has a fixed point, leading to non-Gaussian fluctuations close to $j=0$, is also extended and shifted relative to $F(\theta)$; see Fig.~\ref{figdp1}. This fixed-point region is studied in more detail in Sec.~\ref{seclownoise} to obtain $I(j)$ as $\sigma\ra 0$.

\subsection{Fluctuation relation and upper bounds}

The SCGF and rate function of the current are constrained by the general symmetry of the entropy production,
\be
\frac{P(\Sigma_T=s)}{P(\Sigma_T=-s)}=e^{Ts}.
\ee
which implies, by applying the change of variables (\ref{eqep1}) and by neglecting the potential boundary terms \footnote{The potential boundary terms can be neglected here because they are bounded. For Markov processes with unbounded state space, such as diffusions in $\reals^d$, they have to be included in the large deviation analysis as they can scale with $T$.}, 
\be
I(-j)=I(j)+cj
\label{eqfr1}
\ee
or equivalently
\be
\lambda(k)=\lambda(-k-c),
\label{eqfr2}
\ee
where $c=2\gamma/\sigma^2$. These symmetries for the large deviation functions are collectively referred to as \emph{fluctuation relations} \cite{gallavotti1995a,gallavotti1995,kurchan1998,lebowitz1999} (see \cite{harris2007} for a review) and are connected for the entropy production to a general symmetry of its tilted generator; see Sec.~5 of \cite{lebowitz1999}. For the current, this operator symmetry, which takes the form
\be
\cL_k^\dag=\cL_{-k-c},
\label{eqfr3}
\ee
is not a priori satisfied, since $J_T$ is only proportional to $\Sigma_T$ in the limit $T\ra\infty$ because of the potential boundary terms in (\ref{eqep1}). For $V_0=0$, however, $J_T$ is exactly proportional to $\Sigma_T$ for all $T$, so the operator symmetry (\ref{eqfr3}) holds, as can be verified from the expression (\ref{eqtg1}) of $\cL_k$. In this case, the symmetry (\ref{eqfr2}) thus holds, not just for the dominant eigenvalue in fact but for the whole spectrum.

Two upper bounds constraining $\lambda(k)$ and $I(j)$ can also be derived. The first follows by noting, as before, that the auxiliary force is asymptotically given by $F_k(\theta)=\gamma+\sigma^2 k$ as $k\ra\pm\infty$. Inserting this in the variational principle (\ref{eqopt2}) yields
\be
I_\qap(j)=\frac{(j-\gamma)^2}{2\sigma^2} + \frac{V_0^2}{4\sigma^2}
\label{eqdrb1}
\ee
and since this is not the true minimizer of (\ref{eqopt1}) in general, we must have
\be
I(j)\leq I_\qap(j).
\label{eqdrb2}
\ee
The second bound follows from a result recently derived for jump processes in \cite{gingrich2016} (see also \cite{pietzonka2015}), which here takes the form
\be
I(j)\leq I_\ent(j)=\frac{(j-j^*)^2}{4{j^*}^2}\Sigma^*,
\label{eqentb1}
\ee
where $\Sigma^*=2\gamma j^*/\sigma^2$ is the mean entropy production associated with the mean current $j^*=\lambda'(0)$. 

\begin{figure}[t]
\includegraphics{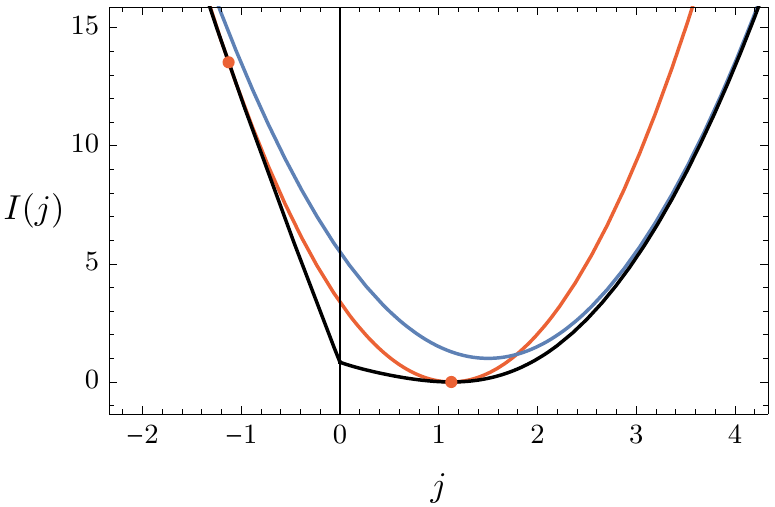}
\caption{(Color online) Auxiliary upper bound (blue curve) and entropic upper bound (red curve) on the rate function $I(j)$ (black curve). The two dots mark the points ($j^*$ and $-j^*$) where the entropic bound touches $I(j)$. Parameters: $V_0=1$, $\gamma=1.5$, $\sigma=0.5$.}
\label{figub1}
\end{figure}

These two bounds are quadratic in $j$ but limit the rate function in different ways, as can be seen in Fig.~\ref{figub1}. The entropic bound (\ref{eqentb1}) is tight at the mean current $j^*$ as well as at $-j^*$, since it satisfies the fluctuation relation (\ref{eqfr1}), but departs from the true $I(j)$ in the tails. By contrast, the bound (\ref{eqdrb2}) obtained from the auxiliary process approximation is tight in the tails but not around the mean by construction. It also gives a non-trivial bound for $\gamma= 0$. The two bounds are identical and in fact equal to $I(j)$ when $V_0=0$, since $I(j)$ is again exactly quadratic with $j^*=\gamma$.

\section{Low-noise limit}
\label{seclownoise}

The exact result (\ref{eqqrft1}) obtained for $V_0=0$ and the numerical results shown in Fig.~\ref{figld2} for $V_0\neq 0$ suggest the following scaling of the rate function:
\be
I(j)\sim\frac{\tI(j)}{\sigma^2}
\ee
as $\sigma\ra 0$. This scaling is also expected from the Freidlin-Wentzell-Graham (FWG) theory of large deviations in the low-noise limit \cite{freidlin1984,graham1989} and implies the following large deviation principle for the current distribution:
\be
P(J_T=j)\approx e^{-T \tI(j)/\sigma^2}
\ee 
in the limit of large $T$ and small $\sigma$. 

Our goal in this section is to obtain the rescaled rate function
\be
\tI(j)=\lim_{\sigma\ra 0}\sigma^2 I(j)
\ee
characterizing the current fluctuations in this limit. Unfortunately, it is not possible to obtain this function analytically or numerically for $\sigma=0$: finding the zero-noise limit of the spectrum of $\cL_k$, which is similar to the semi-classical limit of Schr\"odinger-type operators (see \cite{simon1983b}), is a difficult problem, even more so for non-hermitian operators, and the numerical diagonalization method that we use becomes ill-conditioned for low $\sigma$ relative to $V_0$ \cite{Note1}. However, we can combine the numerical results that we have with analytical results derived from a random walk approximation of the diffusion \cite{lebowitz1999,mehl2008,speck2012} to obtain a good approximation of the current rate function in the low-noise limit. This is done next.

\subsection{Rescaled large deviations}

\begin{figure}[t]
\includegraphics{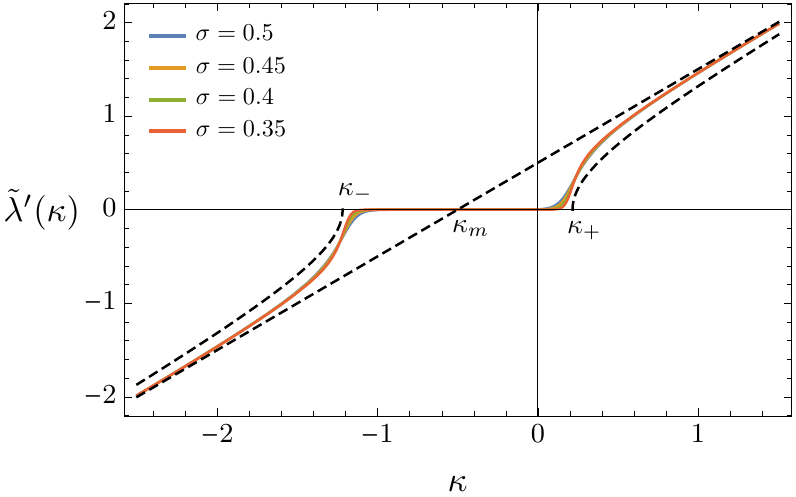}
\caption{(Color online) Derivative of the rescaled SCGF. Parameters: $V_0=1$, $\gamma=0.5$. Dashed lines: linear and hyperbolic approximations.}
\label{figscgf2}
\end{figure}

The rescaled rate function is obtained by rescaling the Legendre transform (\ref{eqlf1}) as
\be
\tI(j)=\sup_{\kappa} \{ \kappa j-\tL(\kappa)\},
\ee
where
\be
\tL(\kappa)=\lim_{\sigma\ra 0}\sigma^2 \lambda\left(\frac{\kappa}{\sigma^2}\right)
\label{eqresc1}
\ee
is the corresponding rescaled SCGF. The result of this scaling is shown for $\tL'(\kappa)$ in Fig.~\ref{figscgf2} and suggests the following properties of $\tL(\kappa)$ whenever $\gamma<V_0$:
\begin{enumerate}
\item\label{p1} Central plateau: $\tL(\kappa)=0$ for $\kappa\in [\kappa_-,\kappa_+]$;

\item\label{p2} Parabolic branches: Far away from the plateau,
\be
\tL'(\kappa)\sim \kappa-\kappa_\md,
\ee 
where $\kappa_\md=(\kappa_++\kappa_-)/2$, implying
\be
\tL(\kappa)\sim \frac{(\kappa-\kappa_\md)^2}{2}
\ee
for $\kappa\gg \kappa_+$ and $\kappa\ll \kappa_-$;

\item\label{p3} Crossover regions: $\tL'(\kappa)$ approaches $\kappa_+$ and $\kappa_-$ continuously, which implies that $\tL(\kappa)$ is continuous with continuous derivatives at these points.
\end{enumerate}

Property \ref{p1} is consistent with the random walk \cite{mehl2008} and FWG approximations \cite{speck2012} of the rate function close to $j=0$ which, when transposed to $\tI(j)$, predict that $\tI(j)$ has a genuine cusp at $j=0$ with left- and right-derivatives given by 
\be
\tI(0^-)=-\frac{\Delta U_-}{\pi}=\frac{1}{\pi}\int_{\theta_s+2\pi}^{\theta_u} F(\theta)d\theta
\label{equ1}
\ee
and 
\be
\tI(0^+)=\frac{\Delta U_+}{\pi}=-\frac{1}{\pi}\int_{\theta_s}^{\theta_u} F(\theta)d\theta,
\label{equ2}
\ee
respectively. Here, $\theta_s$ and $\theta_u$ are the stable and unstable fixed points of $F(\theta)$, respectively, so that $\Delta U_-$ and $\Delta U_+$ are interpreted as the potential barriers created by $F(\theta)$ on the left and the right of $\theta_s$, respectively. By Legendre transform, $\tL(\kappa)$ must then have a genuine plateau for $\kappa\in [\kappa_-,\kappa_+]$ with $\kappa_-$ and $\kappa_+$ equal to the derivatives above. This is confirmed by our numerical results which show that the range $[k_-,k_+]$ where the non-scaled $\lambda(k)$ appears to have a flat plateau grows with $\sigma$ according to
\be
\sigma^2 k_\pm= \kappa_\pm+O(\sigma^2).
\ee
Rescaling $\lambda(k)$ using (\ref{eqresc1}) then implies that $\tL(\kappa)$ has a fixed plateau over $[\kappa_-,\kappa_+]$ that does not scale with $\sigma$. Property \ref{p2} follows from the same rescaling by noting that $\lambda(k)\sim \sigma^2 k^2/2$ for large $|k|$. 

\begin{figure}[t]
\includegraphics{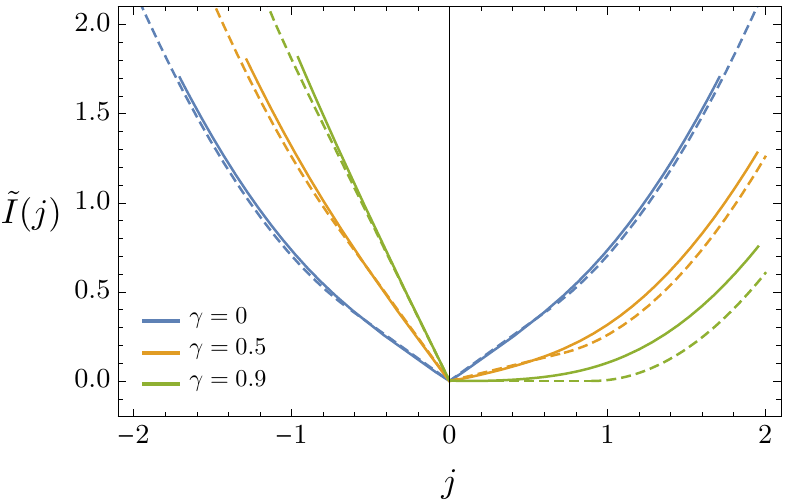}
\caption{(Color online) Full lines: Rescaled rate function obtained with $\sigma=0.35$. Dashed lines: Piecewise-parabolic approximation. Other parameter: $V_0=1$.}
\label{figrfct2}
\end{figure}

These results for $\tL(\kappa)$ imply, as already mentioned, that $\tI(j)$ has a cusp at $j=0$ and parabolic branches. This is verified in Fig.~\ref{figrfct2} which shows $\tI(j)$ as obtained by rescaling $I(j)$ for $\sigma=0.35$, the lowest $\sigma$ that is stable numerically. The cusp is clearly visible and has slopes matching the expected values $\kappa_-$ and $\kappa_+$. Moreover, it can be verified from (\ref{equ1}) and (\ref{equ2}) that
\be
\kappa_+ +\kappa_-=-2\gamma,
\label{eqfr4}
\ee
in agreement with the fluctuation relation (\ref{eqfr1}) rescaled to $\tI(j)$. When $\gamma=0$, for example,
\be
\kappa_+=-\kappa_-=\frac{2}{\pi}, 
\ee
whereas for $\gamma=V_0$, $\kappa_+=0$ and $\kappa_-=-2\gamma$. In all cases, $\kappa_\md=-\gamma$.

In the crossover regions close to $\kappa_-$ and $\kappa_+$, we are not able to determine the exact behavior of $\tL(\kappa)$. As a simple approximation, we can write
\be
\tL'(\kappa)\approx
\left\{ 
\begin{array}{lll}
\kappa-\kappa_\md & & \kappa\notin [\kappa_-,\kappa_+]\\
0 & & \kappa\in [\kappa_-,\kappa_+]
\end{array}
\right.
\label{eqapprox1}
\ee
leading to 
\be
\tL(\kappa)\approx
\left\{ 
\begin{array}{lll}
\displaystyle\frac{(\kappa-\kappa_\md)^2}{2}-\frac{\bu_+^2}{2} & & \kappa>\kappa_+\\
0 & & \kappa\in [\kappa_-,\kappa_+]\\
\displaystyle\frac{(\kappa-\kappa_\md)^2}{2}-\frac{\bu_-^2}{2} & & \kappa<\kappa_-,
\end{array}
\right.
\ee
where
\be
\bu_\pm = \kappa_\pm-\kappa_\md= \frac{\kappa_\pm - \kappa_\mp}{2}.
\ee
The rate function obtained from this piecewise-parabolic approximation shows a fairly good agreement with $\tI(j)$ for $\gamma\approx 0$, as shown in Fig.~\ref{figrfct2}. A better approximation is obtained for $\gamma\lesssim V_0$ by replacing the derivative jumps in (\ref{eqapprox1}) by hyperbolic branches, similar to the deterministic case (see Fig.~\ref{figmean1}), starting at $\kappa_-$ and $\kappa_+$. Both approximations recover the cusp of $\tI(j)$ at $j=0$ and also satisfy the fluctuation relation.

These results apply for $\gamma<V_0$, which corresponds to the fixed-point regime where $j^*=0$ in the low-noise limit. For $\gamma \geq V_0$, the values of $\kappa_-$ and $\kappa_+$ are no longer given by (\ref{equ1}) and (\ref{equ2}), since there are no fixed points in the running regime. The random walk approximation is also not applicable anymore, but our numerical results still suggest that $\tL'(\kappa)$ has the generic form shown in Fig.~\ref{figscgf2} with $\kappa_-$ and $\kappa_+$ satisfying (\ref{eqfr4}) so that $\kappa_\md=-\gamma$. The rate function therefore also have in this case a cusp at $j=0$ with left and right slopes given by $\kappa_-$ and $\kappa_+$, respectively, and is a parabola for large and small current values, which does not depend on $V_0$ as before. The potential, as in the case $\gamma<V_0$, only determines the region of $I(j)$ near the cusp.

\subsection{Auxiliary process and instantons}

\begin{figure}[t]
\includegraphics{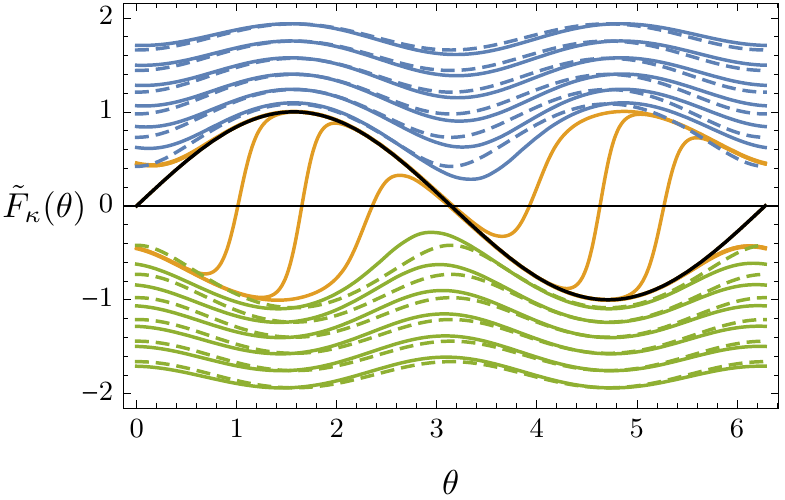}
\caption{(Color online) Full curves: Effective force $\tF_\kappa(\theta)$ of the auxiliary process after rescaling $k$ (see Fig.~\ref{figdp1} for the color code). Dashed curves: Instanton force $G_j(\theta)$ obtained from the FWG calculation. Parameters: $\gamma=0$, $V_0=1$, $\sigma=0.4$.}
\label{figub2}
\end{figure}

We conclude our study of the driven periodic diffusion by comparing the low-noise instanton solution obtained from the FWG theory and the force of the auxiliary process rescaled with $\kappa=\sigma^2 k$. We do not repeat the FWG calculation; see Sec.~3.4 of \cite{speck2012}. The basic idea is that $\tI(j)$ is determined by a deterministic trajectory, called the most probable path or instanton, which minimizes the path action
\be
\cI_T[\theta]=\frac{1}{2T}\int_0^T [\dot\theta_t-F(\theta_t)]^2\, dt
\ee
subject to the constraint $J_T=j$. Thus
\be
\tI(j)=\lim_{T\ra\infty}\min_{\theta_t: J_T=j} \cI_T[\theta]=\cI_\infty[\theta^*],
\label{eqfw1}
\ee
where $\{\theta_t^*\}_{t=0}^T$ is the instanton solving the constrained optimization problem.

We plot in Fig.~\ref{figub2} the time-derivative of $\theta_t^*$ as a function of $\theta_t^*$ (modulo $2\pi$) to obtain the dynamics of the instanton as 
\be
\dot\theta^*_t=G_j(\theta^*_t)
\label{eqfw2}
\ee
for different values of the constraint $J_T=j$ and compare the effective force $G_j(\theta)$ \footnote{The effective potential mentioned in \cite{speck2012} is not what we call the effective force $G_j$ for the instanton or the effective force $F_k$ for the auxiliary dynamics.} thus obtained for the instanton with the effective force of the auxiliary process rescaled as 
\be
\tF_\kappa (\theta) = \lim_{\sigma\ra 0} F_{k=\kappa/\sigma^2}(\theta),
\ee
to follow the rescaling (\ref{eqresc1}) of the SCGF. The results for $G_j$ and $\tF_\kappa$ are in good agreement, given that we can only calculate the later numerically for $\sigma$ no smaller than $0.4$, and are consistent with the idea that the instanton dynamics corresponds to the $\sigma\ra 0$ limit of the auxiliary process \cite{chetrite2014}. In that limit, the SDE (\ref{eqauxp1}) does indeed converge to the following differential equation:
\be
\dot\theta_t=\tF_\kappa(\theta_t),
\label{eqins1}
\ee
which must coincide with (\ref{eqfw2}), after matching $\kappa$ to the current fluctuation $j$ via $\tL'(\kappa)=j$, in order for the optimization problem (\ref{eqopt2}) to be consistent with the FWG optimization (\ref{eqfw1}). The former optimization is solved by the auxiliary dynamics while the latter is solved by the instanton dynamics. Since both give the same rate function, they must therefore describe  the same deterministic dynamics.

With this correspondence, we can understand the dynamical phase transition associated with the cusp of $\tI(j)$ by noting that, although the effective force $\tF_\kappa$ of the auxiliary process is continuous in $\kappa$, it does not create any current for $\kappa\in[\kappa_-,\kappa_+]$, in agreement with $\tL'(\kappa)=0$, since it has a stable fixed point for these values of $\kappa$ (orange lines in Fig.~\ref{figub2}) that prevents rotation as $\sigma\ra 0$. Only when $\kappa<\kappa_-$ or $\kappa>\kappa_+$ does the fixed point disappear and the dynamics become free to rotate (without noise) to create a negative or positive current (blue and green lines in Fig.~\ref{figub2}), determined in the auxiliary dynamics by $\tL'(\kappa)=j$. The cusp in $\tI(j)$ appears therefore as a result of a ``switching'' between two solution or dynamics, as is common in first-order or discontinuous phase transitions \cite{bunin2013,baek2015,bertini2010}. At the switching point, there is an infinite number of dynamics or solutions determined by $\kappa\in [\kappa_-,\kappa_+]$ that produce no current. This was noticed in \cite{mehl2008} and is seen for a different random walk model \cite{gingrich2014}. Some features of the dynamical phase transition reported here are also seen in the 1D periodic WASEP model \cite{espigares2013}.


\section{Conclusion}

We have studied in this paper the current fluctuations of a periodic driven diffusion, often used as a simple model of nonequilibrium systems. Complementing previous studies on this model, we have obtained the large deviation functions of the current, describing its fluctuations in the long-time limit, and have also determined the auxiliary process that explains how these fluctuations are created by an effective noise-induced force. This auxiliary process is useful, as we have demonstrated, for understanding the different fluctuation regimes arising in this model, for deriving bounds on large deviation functions, as well as for deriving the low-noise behavior of the model. Our results, for the latter point, show that the FWG theory of instantons can be obtained by taking the zero-noise limit of the auxiliary dynamics. This is potentially useful for deriving low-noise large deviations of other models, including many-particle models studied within the macroscopic fluctuation theory using the same concepts of most probable paths, instantons, and stochastic action.

For future work, it would be interesting to see whether the numerical results obtained here for the low-noise limit can be obtained analytically from the optimization problems (\ref{eqopt1}) or (\ref{eqopt2}). The existence of a cusp in the rate function is guaranteed by the random walk approximation of the model, but the precise shape of the rate function around the cusp is yet to be determined analytically. Another interesting problem is to see whether the entropic bound has any interpretation in terms of the auxiliary process. Any process that is not the auxiliary process yields, using the optimization problem (\ref{eqopt2}),  an upper bound on the rate function, so it is natural  to look again at this problem to understand large deviation bounds.

\begin{acknowledgments}
We thank Krzysztof Gawedzki for sharing his low-noise approximation of the current rate function, as well as Rosemary J. Harris and Carlos P\'erez-Espigares for useful comments on this paper. P.T.N.\ is supported by a DAAD Scholarship. H.T.\ is supported by the  National Research Foundation of South Africa (Grants no.\ 90322 and no.\ 96199) and Stellenbosch University (Project Funding for New Appointee).
\end{acknowledgments}

\bibliography{masterbib}

\end{document}